\documentclass[11pt,a4paper]{article}

\usepackage[T1]{fontenc}
\usepackage[utf8]{inputenc} 
\usepackage{graphicx}
\usepackage{amsmath,amssymb}
\usepackage{cite}
\usepackage[colorlinks=true,linkcolor=blue,citecolor=blue,urlcolor=blue]{hyperref}
\usepackage{orcidlink}
\usepackage{QCDEX2026}

\begin{document}

\IfFileExists{QCDEX2026-Logo.png}{\noindent\includegraphics[width=0.4\textwidth]{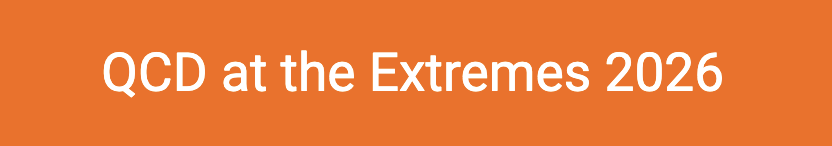}\par\vspace{0.6em}}{}
\begin{center}
{\LARGE\bfseries Kaon isospin-symmetry difference in $pp$ and $AA$ collisions\par}
\vspace{0.8em}
{\large A.V.~Guskov
\quad and\quad
G.I.~Lykasov$^*$\orcidlink{0000-0002-1544-6959}
\quad and\quad 
A.I.~Malakhov
\quad and\quad 
A.A.~Zaitsev$^*$\orcidlink{0000-0003-4711-9925}\par}
\vspace{0.5em}
{\small Joint Institute for Nuclear Research, 141980 Dubna, Moscow region, Russia\par}
\vspace{0.35em}
{\small $^*$Corresponding author: \href{mailto:lykasov@jinr.ru}{lykasov@jinr.ru}\par}
\vspace{0.35em}
{\small\itshape refereed by the conveners of QCDEX2026\par}
\end{center}
\vspace{0.5em}

\begin{abstract}
\setlength{\emergencystretch}{2em}
The energy dependence of the charged-to-neutral kaon production ratio,
$R_K=(\sigma_{K^+}+\sigma_{K^-})/(2\sigma_{K^0_S})$, is investigated in $pp$ and $AA$ collisions. In $pp$ collisions, $R_K$ decreases with increasing centre-of-mass energy $\sqrt{s}$ from threshold to about 20--30 GeV and approaches unity at higher energies. Calculations based on the BMLZ similarity approach, which incorporates the gluon transverse-momentum-dependent distribution at low QCD scales, and the Lund model (PYTHIA) are compatible with the NA61/SHINE $pp$ data. Similar calculations for Ar+Sc (BMLZ) and Pb+Pb (PYTHIA) exhibit comparable energy dependences. Many experimental results for nuclear collisions lie above unity, although their uncertainties vary substantially among experiments; the NA61/SHINE Ar+Sc measurement is considerably more precise than many of the older measurements. The discrepancy between the calculations and the nuclear-collision data may reflect nuclear effects not included in the calculations or possible isospin-symmetry violation. Further precise measurements and theoretical studies are needed to distinguish these possibilities.
\end{abstract}


\vspace{1.0cm}

		The concept of isotopic spin, or isospin, was introduced in \cite{Heisenberg:1932,Wigner:1937} to describe the approximate symmetry between protons and neutrons. It has also been applied to collective nuclear states \cite{Werner:2006}. In the isospin-symmetric limit, the strong-interaction Hamiltonian is invariant under isospin transformations. An analogous symmetry applies to systems containing $u$ and $d$ quarks. However, hadron production in $NN$ and $AA$ collisions depends on the isospin composition of the initial state and on the dynamics of hadronization.

The fragmentation of $u$ and $d$ quarks into kaons depends on the collision energy and kaon species. Consequently, the ratio $R_K=(\sigma_{K^+}+\sigma_{K^-})/(\sigma_{K^0}+\sigma_{\bar K^0})$ depends on production dynamics. A deviation from unity alone is therefore not sufficient evidence of isospin-symmetry violation without accounting for the initial-state isospin composition and relevant production mechanisms.

Possible isospin-symmetry violation in kaon production in nuclear collisions has recently been discussed in \cite{Isotp_vial}. The measured ratio exceeds unity over a broad energy range and is consistent with unity near $\sqrt{s}=3$ TeV. The NA61/SHINE and world data are discussed in \cite{Gazdzinski:2026}. Phenomenological estimates and calculations within UrQMD and the hadron resonance gas (HRG) model are presented in \cite{Gazdzinski:2026,Stepaniak:2023,Giacosa:2025}. These calculations yield $R_K<1$ at some energies below 5--6 GeV, possibly reflecting contributions from meson resonances.

A compilation of $R_K$ data from $e^+e^-$ collisions and electron--deuteron deep inelastic scattering is also presented in \cite{Gazdzinski:2026}. The reported ratios range approximately from 1 to 1.5, with substantial experimental uncertainties. Their interpretation requires consideration of the corresponding interaction dynamics.

In this paper, we investigate the role of kaon production dynamics in the energy dependence of $R_K$ in $pp$ and $AA$ collisions.

We shall follow the similarity approach
	proposed in \cite{4,5} to hadron production in $AA$
	collisions at small transverse momenta.
	It is modified in \cite{LM:2018,ML:2020,LMZ:2021,LMZ:2024}. The conservation law of four-momenta
	is the following:
	
	\begin{equation}
		{(N_AP_A + N_BP_{B} - p_1)}^2 = 
		{(N_Am_0 + N_B m_0 + M)}^2 ,
		\label{eq:n2}
	\end{equation}
          where $N_A$ and $N_B$ are the minimum fractions of the four-momenta $P_A$
        and $P_B$ of nuclei $A$ and $B$ at which the production of a particle of mass $m_1$
        with four-momentum $p_1$ is possible; $m_0$ is the mass of the nucleon; $M$ is the
        effective mass parameter ensuring conservation of the baryon number, strangeness
        and other quantum numbers.
        For symmetric collisions at midrapidity $N_A=N_B=N$.
	This allows us to determine the minimal value of $M$.   
	For $\pi$-mesons $m_1 = m_\pi$  and $M = $0.
	For antinuclei $M=m_1$ and for $K^-$-mesons  $M = m_1 = m_K$,
	$m_K$ is the mass of the $K$-meson.
          For nuclear fragments
          in Eq.~\ref{eq:n2} $m_1$ needs to be subtracted, as well as in Eq.~\ref{eq:n10},  
          since this mass is already present in the original nuclei. 
	For $K^+$-mesons $m_1 = m_K$ and $M = m_\Lambda  - m_0$,
	$m_\Lambda$ is the mass of the $\Lambda$-baryon.
	
	The inclusive spectrum of hadron $h$ produced in the $AB$ collision
	can be parameterized as a general universal function dependent on the 
	similarity parameter $\Pi$, as it was shown in \cite{Baldin_AA:1996}:
	\begin{equation}
	  E d^3 \sigma_{AB}/d^3p~=~A_A^{\alpha(N_A)}\cdot A_B^{\alpha(N_{B})}
          \cdot F(\Pi)
		\label{eq:n4} 
	\end{equation}
	where $\alpha(N_A)=1/3 + N_A/3$, $\alpha(N_B)=1/3 + N_B/3$ 
	and the function $F(\Pi)$ is the inclusive spectrum of hadron production in
        the $NN$
	collision. Here $F(\Pi)$ at $y=$ 0 has the same form as the inclusive
        spectrum
	$\rho_{NN}(y=0,p_t)$ \cite{BGLP:2012,GLLZ:2013},
          where it has been presented in two parts: the quark contribution
          $\rho_{q}(y=0,p_T)$ and the gluon contribution to the inclusive spectrum of hadrons
        produced in $NN$ collisions:
          
	\begin{equation}
		\rho_{NN}(y=0,p_T)~=~\rho_{q}(y=0,p_T)~+~\rho_{g}(y=0,p_T)~,
		\label{def:rhoNN}  
	\end{equation}
	with substitution of transverse momentum $p_T$ by $\Pi$
	\cite{LM:2018,LMZ:2024}.
        The quark contribution $\rho_q$ was calculated within the quark-gluon string model (QGSM) \cite{Kaid1,GLLZ:2013,LLM:2023}. The gluon contribution $\rho_g$ was calculated using the unintegrated gluon distribution, also referred to as the gluon transverse-momentum-dependent (TMD) distribution \cite{GLLZ:2013,LLM:2023}. At low QCD scales, this TMD provides the initial condition for QCD evolution within the $k_T$-factorization approach.

The function $\Pi$ has the following form \cite{4,5},
	\begin{equation} 
		\Pi=\min \frac{1}{2} \sqrt{(u_A N_A + u_B N_B)^2}  ,
		\label{eq:Pi} 
	\end{equation} 
	where $u_A$ and $u_B$ are the four-velocities of nuclei $A$ and $B$, $\Pi$ 
	is found from 
	the minimization of Eq.~\ref{eq:Pi} by solving the equation
	\cite{4,5} at $y=$ 0 and $N_A=N_B=N$:                                            
	\begin{eqnarray}
		\frac{d\Pi}{d N}=0
		\label{def:minimPi} 
	\end{eqnarray}  
	The exact solution of Eq.~\ref{def:minimPi} at $y=0$
        has been derived in \cite{4,LM:2018} and it is presented in
        the following form:  
          
	\begin{eqnarray}
		N=\frac{\Pi}{\mbox{cosh}(Y)}\equiv\frac{2m_0\Pi}{\sqrt{s}}, 
		\label{def:N} 
	\end{eqnarray}  
	Here $Y$ is the rapidity of the colliding nuclei; see \cite{4,5,LM:2018} for details.

Therefore, $\alpha(N)=1/3 + 2m_0\Pi/(3\sqrt{s})$.
	Function $F(\Pi)$ has the following form \cite{LM:2018}:
	\begin{eqnarray}
		F(\Pi)=\bigg[ A_q \mbox{exp}\Big(-\frac{\Pi}{C_q}\Big) \\
		\nonumber
		+
		A_g\sqrt{p_T}\phi_1(s) \mbox{exp}\Big(-\frac{\Pi}{C_g}\Big)\bigg] \sigma_{tot}
		\label{def:F} 
	\end{eqnarray}
	where 
	\begin{eqnarray}
		\Pi(s,m_{1T},y)~=~\left\{\frac{m_{1T}}{2m_0\delta_h}+
		\frac{M}{\sqrt{s}\delta_h}\right\}\mbox{cosh}(y)G,
		\label{eq:n10}
	\end{eqnarray}
	\begin{eqnarray}
		G~=~\left\{1+\sqrt{1+\frac{M^2-m_1^2}
			{(m_{1T}+2Mm_0/\sqrt{s})^2\mbox{cosh}^2(y)}\delta_h}\right\}~
		\label{def:G}
	\end{eqnarray}
	Here 
	$\phi_1(s)~=~1-\sigma_{nd}(s)/\sigma_{tot}(s)$, see \cite{LM:2018,LMZ:2021},
	$\delta_h=\left(1 - \frac{s_{th}^h}{s} \right)$;
	$s_{th}^{\pi}\simeq 4m_0^2$;
	$s_{th}^{K^+}=\left(m_0 + m_K + m_\Lambda \right)^2$;
	$s_{th}^{K^-}=(2m_0+2m_K)^2$;
	$M = m_\Lambda - m_0; m_\Lambda = $ 1.115 GeV; 
	$m_K = $ 0.494 GeV; $s_0 = $ 1 GeV$^2$ is introduced to
	make $s/s_0$ dimensionless; $m_0 = 0.938$ GeV;
	$p_{1T}$ and $m_{1T}$ are the transverse momentum and transverse mass
	of the produced hadron $1$; 
	$\sigma_{nd} = (\sigma_{tot} - \sigma_{el} - \sigma_{SD})$ is the 
	non-diffractive cross-section;
	$\sigma_{tot},\sigma_{SD}$ and $\sigma_{el}$ are the total
	cross-section, the single diffractive cross-section and the elastic
	cross-section of $pp$ collisions, respectively.
	See details in \cite{LM:2018,LMZ:2021,LMZ:2024}. 
	\par
\begin{figure*}[htbp]
\centering
	  \includegraphics[width=0.48\linewidth]{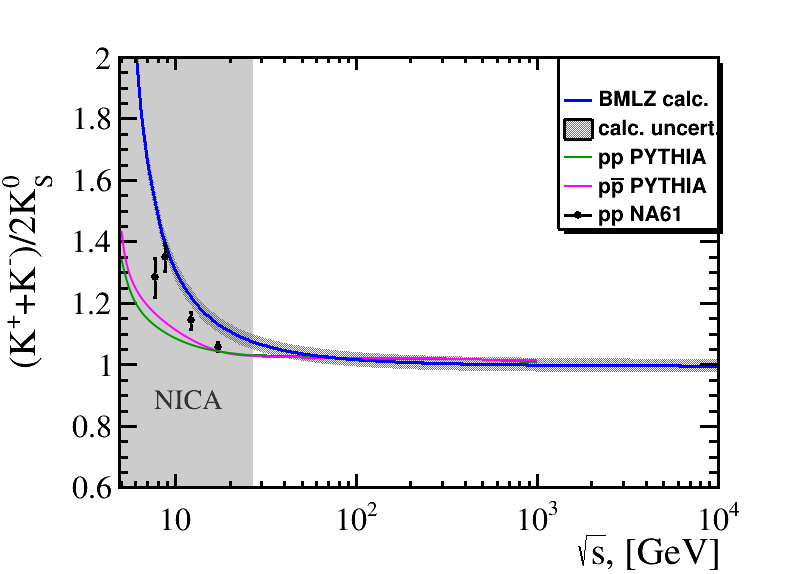}%
          \includegraphics[width=0.48\linewidth]{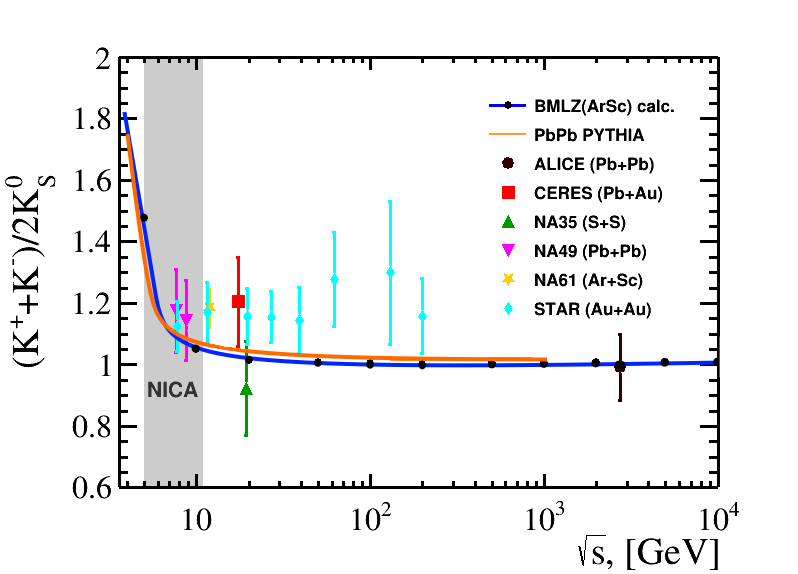}
                    \centering
\caption{\label{fig1} Left: The ratio
				$R_K~=~\frac{\sigma_{K^+}+\sigma_{K^-}}{2\sigma_{K^0_S}}$ for
				$pp$ and $p{\bar p}$ collisions, 
				as a function of $\sqrt{s}$ calculated within the similarity
				approach termed BMLZ \cite{LMZ:2024} (blue line) and the Lund
				model \cite{Lund}
				(green-$pp$, purple-$p{\bar p}$). The band in the blue line
				corresponds to the uncertainties of $R_K$ calculation for $pp$
				collisions. 
				The experimental data are taken from \cite{NA61/SHINE:2017,NA61-K0s,NA61-K0S-2}.
                                Right: The corresponding ratio calculated within the similarity
				approach termed BMLZ for $ArSc$ collisions \cite{LMZ:2024} (blue line) and the Lund
				model \cite{Lund}
				(orange line for $PbPb$). 
				The experimental data are taken from \cite{Isotp_vial}
			}
		\end{figure*}
		
		\par
\begin{figure*}[htbp]
\centering
		  \includegraphics[width=0.47\linewidth]{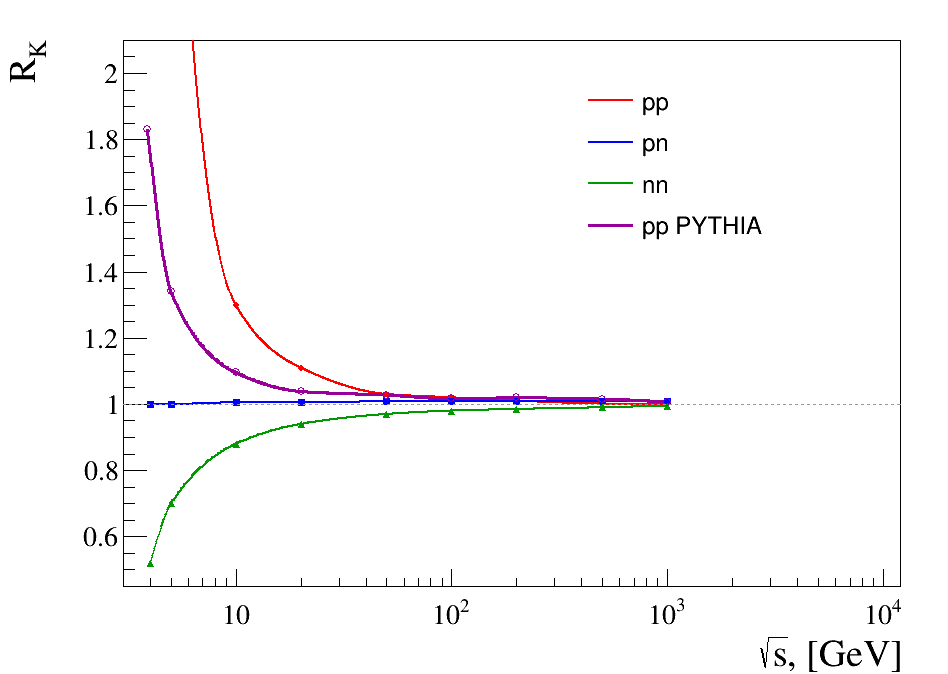}%
                  \includegraphics[width=0.49\linewidth]{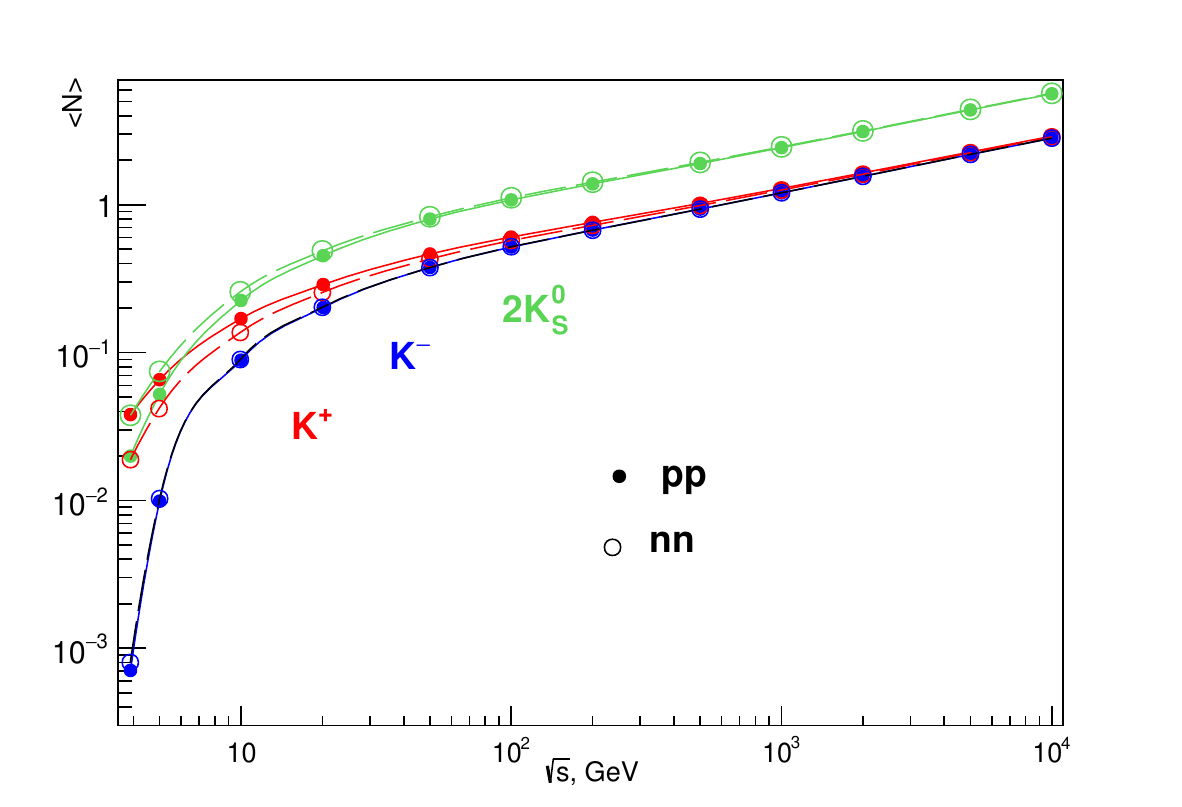}
		\centering
\caption{\label{fig2} Left: The ratio
			$R_K~=~\frac{\sigma_{K^+}+\sigma_{K^-}}{2\sigma_{K^0_S}}$,
			as a function of $\sqrt{s}$ calculated within the similarity
			approach termed BMLZ \cite{LMZ:2024}. The red line corresponds to
			$R_K$ in $pp$, the blue line is $R_K$ in $pn$ and
			the green line is $R_K$ in $nn$ collisions.
                        Right: The dependence of the average multiplicity $\langle N\rangle$ of $K$-mesons on
                                    $\sqrt{s}$
		}
	\end{figure*}

	In Eq.~(7) the first term is the quark contribution and the second term
	is the gluon  contribution to the inclusive
	($NN\rightarrow h X$) spectrum \cite{BGLP:2012,GLLZ:2013}.
	Let us note that the gluon contribution to this inclusive spectrum is
	dominant over the quark contribution at $p_T>$ 0.3 GeV$/$c.
	This dominance is due to the gluon transverse
	momentum distribution (TMD), see details in \cite{LLM:2023,LLM2:2023,LLM3:2024}.

	Using the kaon production cross sections in $pp$ and Ar+Sc collisions calculated according to Eqs.~(\ref{eq:n4}) and (\ref{def:F}), we obtain $R_K$ as a function of $\sqrt{s}$. Figure~\ref{fig1} shows the BMLZ results for $pp$ (left) and Ar+Sc (right), together with Lund-model (PYTHIA) calculations for $pp$, $p\bar p$ (left) and Pb+Pb (right). The calculations are compared with the NA61/SHINE $pp$ data \cite{NA61/SHINE:2017,NA61-K0s,NA61-K0S-2} and the world data for nuclear collisions \cite{Isotp_vial}. The BMLZ and PYTHIA calculations exhibit somewhat different energy dependences below 20 GeV and approach unity at higher energies. The different energy dependences for $pp$, $pn$ and $nn$ collisions shown in the left panel of Fig.~\ref{fig2} help explain why the calculated ratio for Ar+Sc is lower than that for $pp$ at low energies.

Phenomenological studies of the dependence of $R_K$ on the charge-to-mass-number ratio are presented in \cite{Gazdzinski:2026,Stepaniak:2023,Giacosa:2025}. The qualitative expectation of $R_K=1$ for $pn$, $R_K>1$ for $pp$ and $R_K<1$ for $nn$ is consistent with our low-energy calculations in the left panel of Fig.~\ref{fig2}.

Both the Lund and BMLZ approaches account for quark fragmentation into hadrons. Near the kaon production threshold, channels with the lowest allowed final-state masses are particularly important. Examples include associated production, $NN\to N\Lambda K$, and kaon-pair production, $NN\to NNK^+K^-$, with $M=m_\Lambda-m_0$ and $M=m_K$, respectively. The relative contributions of these channels depend on the initial nucleon charges and the available phase space. They can enhance $R_K$ in $pp$ collisions near threshold. As $\sqrt{s}$ increases, additional production channels become accessible and the ratio approaches unity, as shown in Figs.~\ref{fig1} and \ref{fig2}. Nuclear parton distributions are included in the calculations described here, whereas additional nuclear effects, such as rescattering, are not explicitly considered in the interpretation.

Using the inclusive kaon spectra and production cross sections calculated in \cite{LMZ:2021,LMZ:2024}, we have evaluated $R_K$ in $pp$ and Ar+Sc collisions. Similar calculations were performed with PYTHIA \cite{Lund}. For $pp$ collisions, the calculations at $\sqrt{s}=6$--20 GeV are compatible with the NA61/SHINE data and predict a decrease of $R_K$ towards unity.

For nuclear collisions at $\sqrt{s}=6$--10 GeV, the calculations predict a decrease of $R_K$ from approximately 1.35 to 1.05. Several NA49 and STAR measurements lie around 1.15--1.2. At higher energies, several NA61/SHINE, CERES and STAR measurements lie above unity, whereas the NA35 result lies below unity. As shown in the right panel of Fig.~\ref{fig1}, the experimental uncertainties differ substantially among experiments and collision energies. In particular, the NA61/SHINE Ar+Sc measurement has a considerably smaller relative uncertainty than many of the older nuclear-collision measurements. The deviations from unity should therefore be assessed using the uncertainties of each individual measurement rather than a common relative uncertainty. The low-energy predictions can be tested in future measurements at NICA \cite{NICA}.

The energy dependence of the charged-to-neutral kaon production ratio must be interpreted in the context of initial-state isospin composition and kaon production dynamics. Further theoretical and experimental studies are needed to determine whether the observed enhancement in nuclear collisions requires isospin-symmetry violation beyond the effects included in the present calculations.
\section*{Acknowledgments}
		We are very grateful to H.~Jung, M.~Gazdzicki, F~.Giacosa, O.V.~Teryaev for extremely helpful discussions.

\end{document}